\documentclass{aa}
\usepackage{graphicx}
\usepackage[varg]{txfonts}
\usepackage{pdflscape}
\usepackage{amsmath}
\usepackage{empheq}
\usepackage{nccmath}
\usepackage{gensymb}
\usepackage{amssymb}
\usepackage{textcomp}
\usepackage{xfrac}
\usepackage[colorlinks=true, citecolor=blue, linkcolor=blue, urlcolor=blue]{hyperref}
\usepackage{natbib}

\begin{document} 
\title{Deriving physical parameters of unresolved star clusters}
\subtitle{X. Size evolution of young star clusters in the \object{M\,31} disc}
\author{Karolis Daugevi{\v c}ius \and Eimantas Kri{\v s}{\v c}i{\= u}nas \and Rima Stonkut{\.e} \and Vladas Vansevi{\v c}ius}
\institute{Center for Physical Sciences and Technology, Saul\.{e}tekio av. 3, 10257 Vilnius, Lithuania \\\email{rima.stonkute@ftmc.lt}}

\date{Received 30 08 2026; accepted 06 11 2026}

\abstract
{Star cluster structural properties and their correlations with other cluster parameters (age, mass, galactocentric distance) provide key constraints on cluster formation and dynamical evolution within star-forming galactic discs. In the tenth paper of this series, we analysed structural parameters of 1918 young star clusters (up to 800\,Myr) in the disc of \object{M\,31}.}
{We aim to examine the structural evolution of \object{M\,31} disc clusters and determine how their sizes correlate with physical and spatial properties. Our goal is to improve the understanding of cluster formation and evolution in spiral galaxy discs.}
{We use the \object{M\,31} disc cluster parameter catalogue based on photometric measurements of the Panchromatic \textit{Hubble} Andromeda Treasury survey data. We perform statistical analysis to determine how cluster structural parameters relate to their physical characteristics and location within the disc.}
{Young clusters in the disc of \object{M\,31} are compact ($R_{50}\sim1.8$\,pc) and expand rapidly during the first $\sim$20\,Myr, more than doubling in size. At older ages, \object{M\,31} clusters follow a weak age--radius relation ($R_{50}\propto t^{0.12\pm0.01}$). We find no significant correlation between cluster size and galactocentric distance, indicating that internal processes dominate their early evolution, with minimal impact from the galactic tidal field. Our results suggest that stellar-evolution-driven mass loss and internal dynamical interactions are likely drivers of early cluster structural evolution in the \object{M\,31} disc. A shallow mass–radius relation emerges for clusters older than 100\,Myr ($R_{50}\propto M^{0.18\pm0.01}$), while the trend for younger clusters remains uncertain. The diversity of mass–radius relations observed in nearby spiral galaxies, both within and beyond the Local Group, highlights the need for further theoretical work on cluster formation and evolution.}
{}


\maketitle
\nolinenumbers

\section{Introduction}
   \label{Sec1}
   
Since the majority of stars form in clustered environments \citep{Lada2003, PortegiesZwart2010}, star clusters (SCs) provide crucial insight into star formation processes. Their structural properties and internal stellar kinematics encode information about the conditions under which SCs formed and how they subsequently evolve under the influence of internal dynamics and environmental effects \citep{Renaud2018, Krumholz2019, Krause2020}. In the era of the \textit{Gaia} observatory, high-precision astrometry resulted in a significantly expanded census of clusters in the Milky Way (\object{MW}) disc \citep{Cantat-Gaudin2022, Hunt2024}, enabling detailed, star-by-star studies. However, such investigations remain limited to a relatively small volume (radius $\sim$2\,kpc) due to strong interstellar extinction within the Galactic disc \citep{Kharchenko2013, Hunt2024}.

Therefore, to achieve a more complete understanding of cluster formation and evolution across diverse disc environments \citep{Krumholz2019, Adamo2020} – and to better constrain theoretical models \citep{Pfeffer2019, Krause2020, Webb2021} – extragalactic SCs must be studied. Although kinematic data are scarce beyond the \object{MW} and are hampered by the effects of binaries \citep{Gieles2010, Niederhofer2024}, and direct real-time observation of cluster evolution is impossible due to astronomical timescales, statistically large cluster samples allow us to investigate how structural parameters vary and correlate with other cluster properties, such as their age, mass, and location within the host galaxy \citep{Ryon2015, Brown2021, Tang2026}.

Over the past two decades, with the help of the \textit{Hubble} Space Telescope (HST), the cluster age–radius relation ($R_{50}\propto t^{\alpha}$; here $R_{50}$ is half-light, or effective, radius) has been examined extensively in recent studies of spiral galaxies beyond the Local Group \citep{Ryon2015, Ryon2017, Brown2021, Tang2026}, sometimes yielding conflicting results. Notably, \citet{Larsen2004} analysed a combined sample of SCs from 18 nearby spiral galaxies and found no compelling evidence of significant correlation between cluster age and size. In contrast, a more recent study by \citet{Brown2021}, based on young SCs in 31 galaxies from the LEGUS survey \citep{Adamo2017, Cook2019}, reported a statistically significant increase in cluster radius with age. Specifically, they found that between the 1–10\,Myr and 10–100\,Myr age bins, the typical effective radius at a cluster mass $10^4\,M_{\sun}$ grows from 2.36 to 2.51\,pc, while that of clusters with $10^5\,M_{\sun}$ expand from 3.58 to 4.76\,pc. However, no further evolution is observed between the 10–100\,Myr and 0.1–1.0\,Gyr bins.

Considering individual nearby spiral galaxies, the SC population in \object{M\,51} has been studied extensively. Using HST WFPC2+NICMOS3 data, \citet{Bastian2005}, in a cluster sample consisting mostly of clusters younger than 1\,Gyr, detected a weak tendency for younger clusters to be larger than older ones, but found no robust age–radius correlation overall. In contrast, \citet{Lee2005}, based on HST WFPC2 data, reported a weak increase in cluster size with age, with a power-law index of $0.13\pm0.02$ ($0.06\pm0.02$ after accounting for the mass–size relation). Using HST ACS data, \citet{Scheepmaker2007} found that cluster radii increase with colour, particularly in $B-V$, and argued that if the colour range reflects an age spread, this implies expansion with age, consistent with \citet{Lee2005}. Combining WFPC2 and ACS data and applying a similar methodology to \citet{Lee2005}, \citet{Hwang2010} derived a somewhat stronger age–size relation, with a power-law index of $0.25\pm0.03$ ($0.14\pm0.03$ after correcting for the mass–size relation). More recently, \citet{Chandar2016} analysed $\sim$3800 SCs in \object{M\,51} and found stronger evidence for early expansion: clusters more massive than $3\times10^{4} M_{\sun}$ at ages of 100–400\,Myr have $R_{50}$ values larger by a factor of $\sim$(3–4) compared to SCs younger than 10\,Myr. Overall, the evidence increasingly supports cluster size evolution with age in \object{M\,51}.

\object{M\,83} is another nearby spiral with well-studied SC populations. Using observations of two HST WFC3 fields, \citet{Bastian2012} found a significant increase in cluster radius with age. The age–size relation is well described by a power-law with an index of $0.22\pm0.04$ in the inner field and $0.24\pm0.06$ in the outer field. They interpreted this trend as either a physical effect of cluster expansion or a possible selection effect related to the light profile fitting procedure. Expanding the analysis to seven HST WFC3 fields covering a large fraction of the \object{M\,83} disc, \citet{Ryon2015} reported a similar age–radius correlation, best described by a power law with an index of $0.26\pm0.07$. Importantly, they were able to reject the fitting procedure effect proposed by \citet{Bastian2012}, implying that the observed age–radius relation likely reflects genuine cluster size evolution.

\citet{Ryon2017} examined the sizes of young massive SCs ($M\geq5\times10^{3}$ $M_{\sun}$ and $t\leq200$\,Myr) of two LEGUS galaxies, \object{NGC\,628} and \object{NGC\,1313}. In \object{NGC\,1313}, they found the cluster age–size relation to be broadly similar to that reported for \object{M\,83} by \citet{Ryon2015}, with a slope of $\sim$0.3. Meanwhile, the correlation in \object{NGC\,628} is somewhat weaker, with a slope of $\sim$0.1. \citet{Ryon2017} suggest that this difference may be due to observational limitations: the ability to detect smaller clusters in \object{M\,83} and \object{NGC\,1313}, compared to \object{NGC\,628}, provides better leverage for constraining the age–radius relation, leading to stronger inferred correlations.

Another important and widely studied property of cluster populations is the mass–radius relation. For a combined sample of SCs from 18 nearby spirals, \citet{Larsen2004} reported a shallow trend, $R_{50} \propto M^{0.10\pm0.03}$. However, for clusters from the LEGUS survey, \citet{Brown2021} derived a significantly steeper relation, $R_{50} \propto M^{0.24\pm0.01}$. Importantly, they found evidence for a possible evolution of the mass–radius relation with age: the power-law index increases from 0.18$\pm$0.03 for 1–10\,Myr clusters to 0.28$\pm$0.02 for 10–100\,Myr SCs, remaining roughly constant at older ages. However, both studies combined clusters from multiple galaxies, thus these trends may reflect only an average mass–radius relation, potentially masking galaxy-to-galaxy variations.

As in the case of the age–radius relation, studies of the mass–radius correlation in \object{M\,51} have yielded contrasting results. While \citet{Bastian2005} and \citet{Scheepmaker2007} found no evidence for a correlation between cluster mass and radius, \citet{Lee2005} reported a shallow power-law relation, where radius depends on mass with an index of 0.14$\pm$0.03. Similarly, \citet{Hwang2010} derived a best-fit power-law relation with an index of 0.16$\pm$0.02, consistent with the result of \cite{Lee2005}.

For \object{M\,83}, \citet{Bastian2012} found no substantial evidence for a cluster mass–radius relation. In contrast, \citet{Ryon2015} reported a significant correlation, with cluster radius following a power-law of index 0.3$\pm$0.1, and, similarly to \citet{Brown2021}, noted that the relation strengthens with age. \citet{Ryon2017} found no significant relation between cluster mass and radius in \object{NGC\,628}, but in \object{NGC\,1313} they observed a modest cluster radius increase with mass; however, this trend disappeared when the sample was divided into age bins. In a review combining SC data from eight nearby galaxies, \citet{Krumholz2019} concluded that the mass–radius relation is broadly consistent with a general scaling of constant density $R_{50}\propto M^{1/3}$.

Several studies have investigated correlations between cluster radius and galactocentric distance, $\rm R_{GC}$, with mixed results. In \object{M\,51}, \citet{Bastian2005} found no clear relation, whereas \citet{Scheepmaker2007} reported a shallow trend, $R_{50}\propto \rm R_{GC}^{0.12\pm0.02}$. For \object{M\,83}, \citet{Bastian2012} observed slightly larger clusters in the outer field, but attributed this to an age difference; similarly, \citet{Ryon2015} found no significant radius–distance correlation once the age–distance relation was accounted for. No significant trends were detected in \object{NGC\,628} or \object{NGC\,1313} by \citet{Ryon2017}. Within the Local Group, \citet{SanRoman2012} reported no strong relation in \object{M\,33}, but found substantial trends in the \object{MW}, the Andromeda galaxy (\object{M\,31}), and the Magellanic Clouds; however, they were determined based mostly on massive ($>10^{4}\,M_{\sun}$) clusters.

Although several studies have explored cluster size correlations within the Local Group \citep{Vansevicius2009, Gatto2021}, they mostly relied on rather small statistical samples, and there is a lack of homogeneous studies involving large sets of clusters spanning a wide range of parameters. As highlighted above, the most extensive investigations of such relations have instead been carried out in galaxies beyond the Local Group, but these were typically limited to relatively massive clusters \citep[$\geq$$10^{3.5-4.0}~M_{\sun}$;][]{Ryon2015, Ryon2017, Brown2021}. Therefore, detailed studies within the Local Group are needed to probe the low-mass regime and obtain a more complete understanding of cluster formation and evolution in star-forming galactic discs.

Thanks to its proximity and the availability of high-resolution HST imaging from the Panchromatic \textit{Hubble} Andromeda Treasury survey \citep[PHAT;][]{Dalcanton2012, Johnson2015}, \object{M\,31} provides an ideal laboratory for studying low-mass SCs, $\sim$300–30\,000~$M_{\sun}$, within a disc of a spiral galaxy similar to the \object{MW}. The PHAT survey covers more than 20\,kpc of radial distance in the north-eastern part of the \object{M\,31} disc, enabling the investigation of clusters across a wide range of environments and physical properties.

In this study, by employing cluster parameters derived from photometric measurements, we aim to investigate the structural evolution of \object{M\,31} disc SCs and correlations between their sizes and other properties through statistical analysis. Our ultimate goal is to improve the understanding of cluster formation and evolution and the underlying physical processes, while providing constraints for theoretical models by extending the analysis to lower cluster masses than those considered in previous extragalactic star cluster studies.

The paper is structured as follows. In Sec.\,\ref{Sec2}, we describe the cluster parameter catalogue used in this study. In Sec.\,\ref{Sec3}, we examine the structural evolution of clusters with age. In Sec.\,\ref{Sec4}, we investigate the cluster mass–radius relation. Finally, in Sec.\,\ref{Sec5}, we summarise our results and present conclusions.

\section{Star cluster data}
	\label{Sec2}

Physical and structural parameters of SCs in our sample are taken from the catalogue of SCs in \object{M\,31} \citep{Krisciunas2026}. Cluster physical parameters (age, mass, and extinction) were determined using broad-band HST aperture photometry \citep{Naujalis2021, Krisciunas2023}, compared with stochastic stellar population models \citep{deMeulenaer2015, deMeulenaer2017}, as well as through the analysis of colour-magnitude diagrams (CMDs) of resolved stars \citep{Williams2023}. Physical parameters were derived assuming a fixed solar metallicity. In this work, we consider the half-light radius ($R_{50}$; or effective radius), as well as $R_{30}$ and $R_{80}$, which enclose 30\% and 80\% of the total cluster flux, respectively. These structural parameters were determined from growth curves in the ACS/WFC $F475W$ passband, which, in the PHAT survey, offers the best signal-to-noise and is least affected by contamination from resolved field stars of old \object{M\,31} stellar populations. 

Our final sample consists of 1918 SCs younger than 800\,Myr, spanning a wide range of mass, $\sim$300–30\,000\,$M_{\sun}$. Notably, the sample includes a subset of young ($t < 100$\,Myr) and massive ($>$$10^{4}~M_{\sun}$) objects, that are most likely unbound stellar associations or groups rather than genuine bound clusters. These objects are more common among younger stellar populations.

\begin{figure}
	\centering
	\includegraphics[width=9cm]{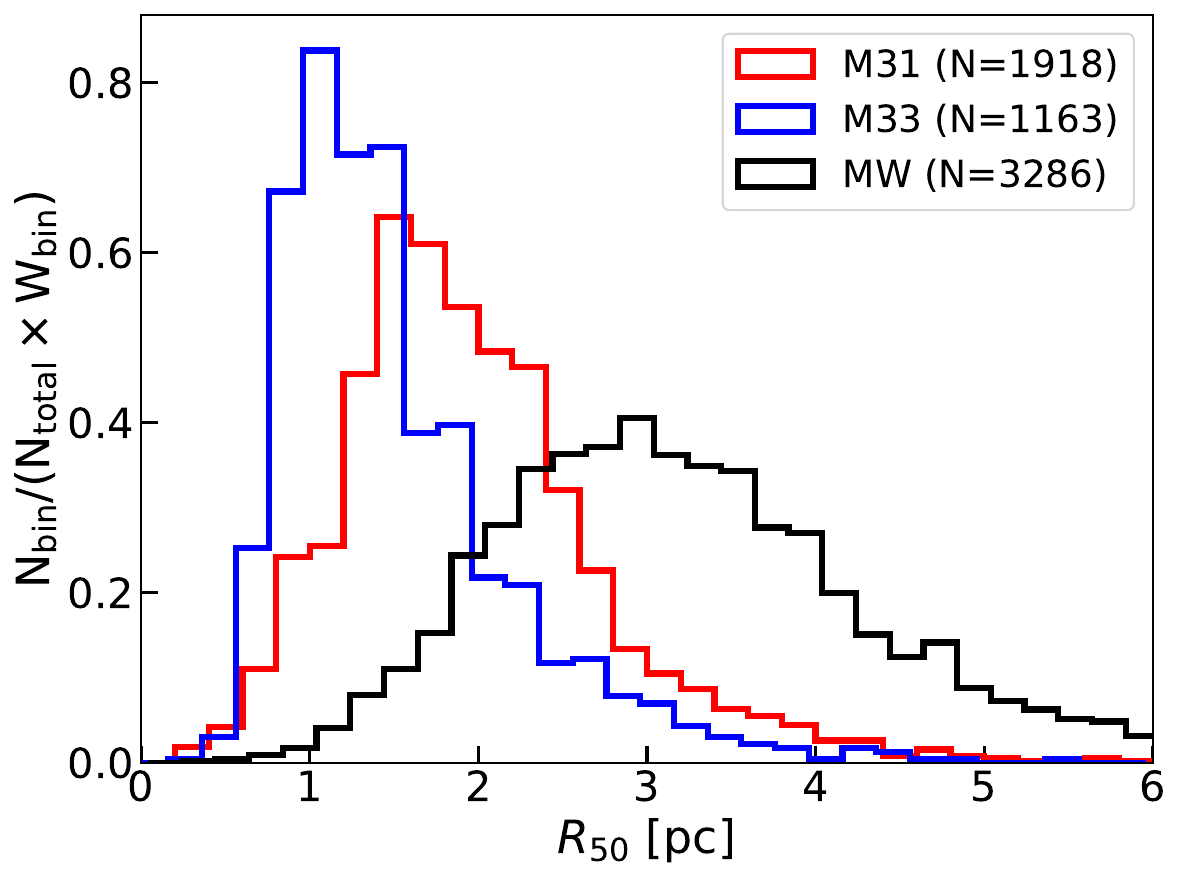}
	\caption{Comparison of star cluster size distribution in \object{M\,31} with those in other spiral galaxies of the Local Group. \object{M\,33} cluster data are taken from \citep{Johnson2022}. \object{MW} cluster data are from \citet{Hunt2024}. For all galaxies, only clusters younger than 800\,Myr are considered. \object{M\,33} and \object{MW} histograms are slightly shifted relative to the histogram of \object{M\,31} for visual clarity.}
	\label{fig1}
\end{figure}

We examine the distribution of cluster $R_{50}$ in the \object{M\,31} disc and compare it with those in other Local Group spiral galaxies (Fig.\,\ref{fig1}). The \object{M\,33} data are taken from the PHATTER SC catalogue \citep{Johnson2022}, where $R_{50}$ values were derived using the same methodology as the one  applied to \object{M\,31} clusters in \citet{Johnson2015}. Our measurements of $R_{50}$ are well correlated with those of \cite{Johnson2015} and show consistent median values, ensuring a robust comparison with \object{M\,33}. In contrast, cluster size estimates for the \object{MW} are taken from the \citet{Hunt2024} \textit{Gaia} SC catalogue, where radii are derived using a different method based on counting cluster member stars. For the \object{MW}, as an analogue of $R_{50}$, we use the on-sky projected radius enclosing 50\% of cluster members within the best-fitting Jacobi radius; detailed definitions are provided in \citet{Hunt2024}. This choice is most appropriate for our analysis, as we aim to compare gravitationally bound clusters. Another size estimate reported in \citet{Hunt2024} -- the radius enclosing 50\% of members including those outside the Jacobi radius --  is influenced by tidal tails, a feature generally not detectable at extragalactic distances, which could possibly introduce inconsistencies in our analysis (E. L. Hunt, priv. comm.). Therefore, the comparison with the \object{MW} cluster sample is less direct and should be treated with caution. We emphasise that we include only those \object{MW} objects that \citet{Hunt2024} classify as ``open clusters''. For both \object{M\,33} and the \object{MW}, we limit the samples to clusters younger than $\sim$800\,Myr and with log$(M/M_{\sun})\leq5.0$ to ensure a consistent comparison with \object{M\,31}.

In Fig.\,\ref{fig1}, we demonstrate that \object{MW} SCs are, on average, more extended than those in \object{M\,31} and \object{M\,33}, while \object{M\,33} clusters are slightly more compact than \object{M\,31} SCs. The $3\sigma$-clipped statistics (median half-light radius, $\tilde{R_{50}}$ with 16th and 84th percentiles) are as follows: $\tilde{R_{50}}=1.82^{+0.76}_{-0.57}$\,pc for \object{M\,31}; $\tilde{R_{50}}=1.35^{+0.66}_{-0.45}$\,pc for \object{M\,33}; $\tilde{R_{50}}=3.10^{+1.19}_{-0.95}$\,pc for \object{MW}. 

The \object{MW} cluster size distribution is consistent with results for other nearby galaxies ($\lesssim$10\,Mpc) outside the Local Group, where the typical peak of young cluster size distributions is $R_{50}\approx3$\,pc \citep{Larsen2004, Lee2005, Scheepmaker2007, Ryon2015, Ryon2017, Brown2021}. On the other hand, SCs in \object{M\,31} and \object{M\,33} appear to be substantially more compact compared to those in other nearby galaxies. These discrepancies may be due to differences in size measurement techniques or selection effects. However, these discrepancies may also reflect an environmental dependence of the characteristic cluster size, especially given that previous studies \citep{Kodaira2004, Vansevicius2009, SanRoman2012} have suggested a compact nature of SCs in \object{M\,31} and \object{M\,33}.

Regarding \object{M\,31}, studies based on Subaru Suprime-Cam ground-based observations of the south-western part of the galaxy's disc \citep{Kodaira2004, Sableviciute2006, Sableviciute2007, Vansevicius2009} have shown that SCs in this region are typically compact. In particular, \citet{Vansevicius2009} report a characteristic $R_{50}\sim1.5$\,pc, in good agreement with the results presented here, despite differences in instrumentation and methodology. Considering that our analysis focuses on the north-eastern part of the disc, this suggests that SCs are compact throughout the entire \object{M\,31} disc. Similarly, a study of 161 SCs in \object{M\,33} based on the HST ACS/WFC observations \citep{SanRoman2009, SanRoman2012} indicates that \object{M\,33} clusters seem to be smaller than those in the \object{MW}, \object{M\,31}, and the Magellanic Clouds. Thus, results presented in Fig.\,\ref{fig1} are consistent with previously reported findings.

\section{Evolution of star cluster structure}
	\label{Sec3} 

\begin{figure*}
	\centering
	\includegraphics[width=18.0cm]{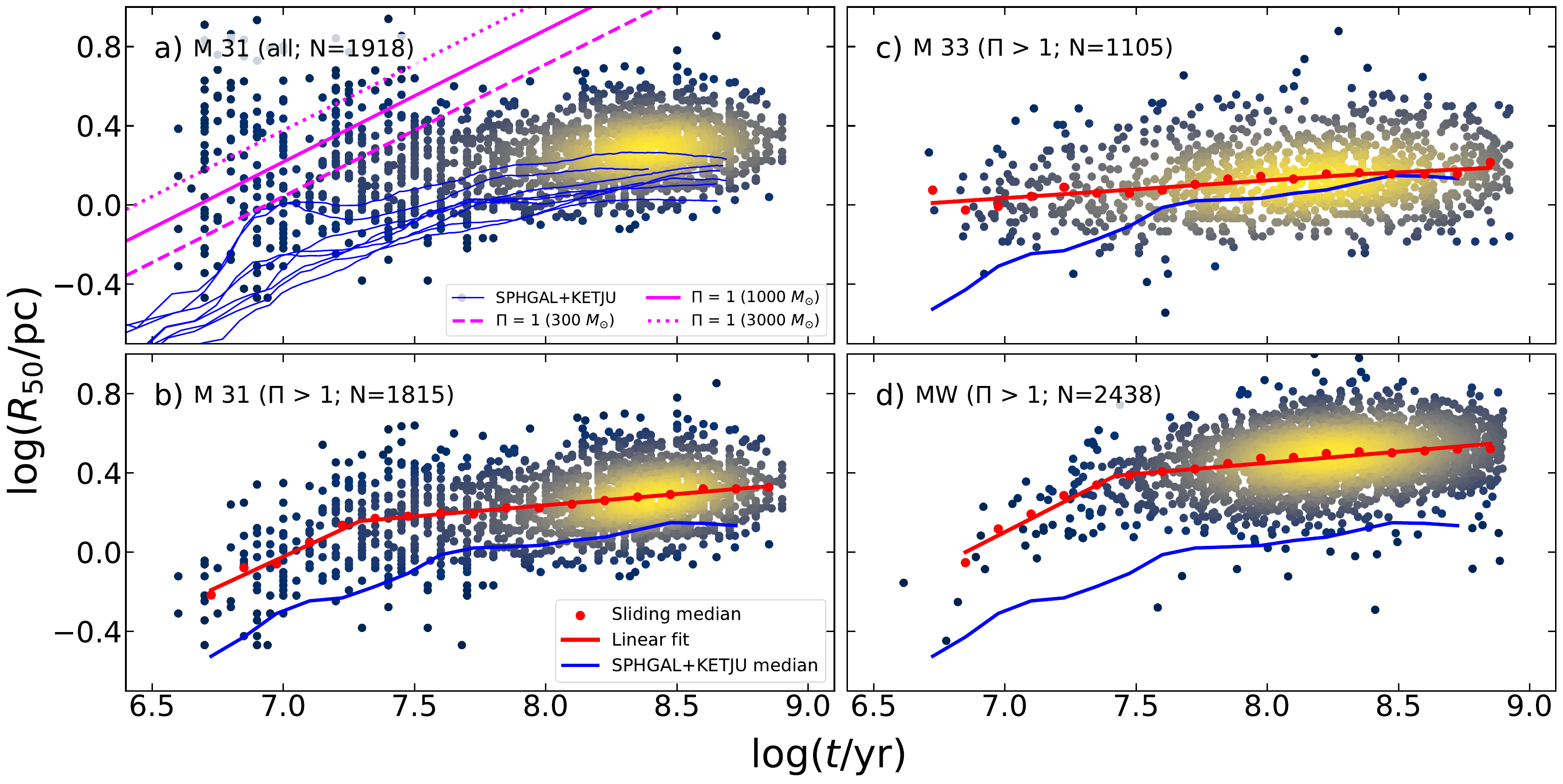}
	\caption{Cluster size evolution with age. Panel a) shows all 1918 \object{M\,31} objects considered in this study; panel b) includes only those \object{M\,31} clusters that satisfy the dynamical age selection criterion $\Pi>1$. Panel c) shows results from the \object{M\,33} star cluster catalogue \citep{Johnson2022}. Panel d) presents results from the \object{MW} \textit{Gaia} cluster catalogue \citet{Hunt2024}; here, only objects classified as ``open clusters'' are included. For \object{M\,33} and \object{MW}, we consider clusters with $t\leq800$\,Myr and log$(M/M_{\sun})\leq5.0$ that satisfy the $\Pi>1$ criterion. In panels b)--d), red markers show running medians of $\log(R_{50}/\mathrm{pc})$, calculated in overlapping 0.25-dex-wide age windows spaced by 0.125\,dex; the same age-bin centres are used in all three panels. Red lines show linear fits to the young- and old-age median points. The young and old fitting ranges, $\log(t/\mathrm{yr})$, are: $\leq$7.35 and $\geq$7.25 in panel b), $\leq$7.85 and $\geq$7.75 in panel c), and $\leq$7.45 and $\geq$7.35 in panel d); each pair of lines is drawn to its mutual intersection. Blue curves in panel a) show evolutionary tracks of $\tt SPHGAL + KETJU$ star cluster models \citep{Lahen2025a}, smoothed using a sliding window of width 0.25\,dex in log$(t/\rm yr)$ axis, while the blue curves in panels b)--d) show the median model evolution. Magenta lines in panel a) indicate the $\Pi=1$ boundary for different cluster masses.}
	\label{fig2}
\end{figure*}

In this section, we examine how cluster structural properties vary with age. Our goal is to evaluate the structural evolution of gravitationally bound SCs. To minimise contamination by unbound systems, we follow the prescription of \citet{Gieles2011}. To distinguish clusters from associations, they introduce the dynamical age, $\Pi$, which is the ratio between cluster age, $t$, and the crossing time of member stars in the cluster, $t_{\rm cr}$:
\begin{fleqn}
	\begin{align}
	\Pi = t / t_{\mathrm{cr}},
	\end{align}
\end{fleqn}
with crossing time defined as follows:
\begin{fleqn}
	\begin{align}
		t_{\mathrm{cr}} \equiv 10 \left( \frac{R_{50}^{3}}{G M} \right)^{1/2},
	\end{align}
\end{fleqn}
where $G$ is the gravitational constant, and $M$ is the cluster's mass. \citet{Gieles2011} suggest that a distinction between bound and unbound objects can be made at $\Pi=1$, where $\Pi>1$ are bound clusters. They note, however, that while the $\Pi=1$ boundary provides a robust distinction between clusters and associations for objects with $t\gtrsim10$\,Myr, it is less reliable at younger ages; nonetheless it still offers a satisfactory separation for our sample of objects.

In Fig.\,\ref{fig2}, we present $R_{50}$ as a function of cluster age. The left panels show results for the \object{M\,31} disc SCs: panel a) includes all 1918 objects, while panel b) shows only those that satisfy the $\Pi>1$ selection criterion. When no limit on $\Pi$ is applied, no clear trend in cluster size growth or contraction is visible. However, the first $\sim$50\,Myr exhibit a significantly larger spread in $R_{50}$, with both more compact and more extended clusters present compared to older populations. This may indicate an ongoing structural evolution, with those less concentrated systems expanding and likely dissolving, while more compact clusters expand but remain gravitationally bound. 

Once the $\Pi>1$ criterion is applied to reject unbound objects, we observe a trend of the median $R_{50}$ increasing with age. The most significant cluster expansion happens within the first $\sim$20\,Myr. In this age interval, when adopting the $\Pi>1$ limit, the median $R_{50}$ increases by more than a factor of two – from $\sim$0.6\,pc to $\sim$1.4\,pc, corresponding to log$(R_{50}/\rm pc)$ from $\sim$(–0.20) to $\sim$0.15. We note that this trend remains evident even when using a less strict ($\Pi>0.5$) threshold. We emphasise that, unless stated otherwise, we restrict the \object{M\,31} cluster sample to objects satisfying the $\Pi>1$ criterion for the remainder of this paper.

At older ages, the median $R_{50}$ continues to increase, albeit at a significantly slower rate, reaching $\sim$2\,pc over the subsequent $\sim$700\,Myr. We quantify the age--radius relation for clusters older than $\sim$20\,Myr, since this sample is largely unaffected by the adopted $\Pi$ selection and can therefore be determined relatively robustly. To derive a slope of the age--radius relation, we divided the cluster sample into contiguous, non-overlapping 0.125-dex age bins (bins of this width reflect variation by age in sufficient detail and still have a statistically significant number of clusters) and calculated the median $\log(R_{50}/\mathrm{pc})$ in each bin. We then fitted these median radii in log--log space, corresponding to a power-law relation $R_{50}\propto t^{\alpha}$. Formal 1$\sigma$ uncertainties on $\alpha$ were derived from the covariance matrix. We obtain a slope of $\alpha=0.12\pm0.01$. 

Compared with the age--size trends discussed in Sec.\,\ref{Sec1}, this value is similar to the slopes reported for \object{M\,51} \citep{Lee2005} and \object{NGC\,628} \citep{Ryon2017}, but it is significantly shallower than those found for \object{M\,83} \citep{Bastian2012, Ryon2015} or \object{NGC\,1313} \citep{Ryon2017}. This weak size evolution may reflect the combined effects of declining stellar-evolution mass loss, slow relaxation-driven expansion, and regulation by the galactic tidal field. In addition, such a shallow trend may be driven, at least in part, by selection bias: as cluster ages increase, only more massive clusters remain detectable, and these are expected to be larger on average than their lower-mass counterparts.

In the right panels of Fig.\,\ref{fig2}, we examine the cluster size–age relation in other Local Group spirals for comparison with our findings in the \object{M\,31} disc. For \object{M\,33} and the \object{MW}, we use the catalogues of \citet{Johnson2022} and \citet{Hunt2024}, respectively. For both cluster samples we apply the $\Pi>1$ selection criterion. Similar to \object{M\,31}, neither galaxy shows strong cluster size evolution beyond the first few tens of Myr, with the median $R_{50}$ remaining roughly constant or increasing only slightly. 

For \object{M\,33}, we do not observe a significant early evolution of $R_{50}$ based on the \citet{Johnson2022} data (Fig.\,\ref{fig2}c). The median $R_{50}$ in \object{M\,33} shows a weak, approximately constant increase across the full age range. This could suggest that the rapid early cluster expansion observed for \object{M\,31} SCs does not occur in \object{M\,33}. However, we suggest that this difference could also arise from methodological effects. Applying the same procedure used above to determine the slope of the age--radius relation, we obtain a power-law index $\alpha=0.06\pm0.01$ for \object{M\,33} clusters older than $\sim$60\,Myr, which is even shallower than the value we derive for \object{M\,31}. Importantly, unlike in \object{M\,31} and the \object{MW}, we do not observe a truncation in the age--radius relation for \object{M\,33}. It can be noted that for the \object{M\,33} clusters younger than $\sim$70\,Myr, we obtain an identical albeit considerably more uncertain slope compared to older clusters -- $\alpha=0.06\pm0.05$.

\citet{SanRoman2012} investigated cluster structural parameters as a function of age for \object{M\,33} SCs, although their sample mainly consists of objects aged from 100\,Myr to 1\,Gyr. They found no correlation between $R_{50}$ and age, in general agreement with the weak trend we determine for \object{M\,31} SCs over the same age range. However, \citet{SanRoman2012} reported that older clusters tend to have a broader range of core radii, $r_{c}$, suggesting possible structural evolution at later ages. Interestingly, \citet{Gatto2021} identified a similar trend for \object{SMC} clusters at ages beyond 200\,Myr.

For the \object{MW}, Fig.\,\ref{fig2}d shows a strong increase in the median cluster radius with age up to $\sim$30\,Myr. However, the number of clusters younger than 10\,Myr is rather low, suggesting that, most likely, it is largely driven by the adopted $\Pi>1$ selection criterion. Therefore, we cannot conclude that the observed trend – which is similar to that seen in \object{M\,31} – reflects a genuine structural evolution of \object{MW} clusters, especially given that the method used to measure the radii for Galactic SCs is fundamentally different, as discussed in Sec.\,\ref{Sec2}. Beyond $\sim$30\,Myr, the \object{MW} cluster population follows an age--radius relation identical to that found for \object{M\,31}, with a slope of $\alpha=0.12\pm0.02$.

In a recent study, \citet{DellaCroce2024} performed a systematic kinematic analysis of young ($t<300$\,Myr) \object{MW} clusters based on \textit{Gaia} DR3 data. They found that the majority of $t\lesssim30$\,Myr SCs are currently undergoing considerable expansion, while older clusters are, in most cases, found to be compatible with the virial equilibrium state. This is consistent with our findings for \object{M\,31}. Notably, \citet{Gatto2021}, who combined their sample of 134 SMC clusters with an additional 51 LMC and SMC clusters studied by \citet{Mackey2003b, Mackey2003a}, based on their core radii measurements, found that SCs younger than 200–300\,Myr are compact and seem to undergo rapid expansion within the first 10–20\,Myr, followed by little structural evolution for the next 200–300\,Myr. \citet{DellaCroce2024} also compared their results in the \object{MW} with N-body simulations of young SCs \citep{Livernois2021} and found the observed expansion during first 30\,Myr is consistent with systems experiencing violent relaxation and evolving towards virial equilibrium. They suggest that residual gas removal and mass loss due to stellar evolution are likely to be key drivers of the observed expansion as well.

In Fig.\,\ref{fig2}, we also compare our results with the cluster evolutionary tracks from the $\tt SPHGAL + KETJU$ simulations, which follow star cluster formation within a star-by-star galaxy model and include hydrodynamics, star formation, stellar evolution, and collisional gravitational interactions between stars and stellar remnants \citep{Lahen2025a}. In particular, we use results from the simulations of a dwarf galaxy (metallicity -- 1\% of solar; virial mass -- $4\times10^{10}\,M_{\sun}$). Despite the large difference between the simulation setup and \object{M\,31}, we find that the model evolutionary tracks exhibit a similar shape to the median $R_{50}$ evolution of the $\Pi>1$ sample, defined by rapid early expansion followed by significantly slower subsequent growth. However, \object{M\,31} clusters are generally larger than simulated ones, thus the agreement is rather qualitative than quantitative.

The offset in cluster size between \object{M\,31} SCs and $\tt SPHGAL + KETJU$ tracks is most likely a cluster mass effect – all simulated clusters have masses $\lesssim$800\,$M_{\sun}$, whereas the majority of SCs in our sample are $\gtrsim$1000\,$M_{\sun}$. Other possible factors for such offset might include metallicity effects, dwarf galaxy environment, and how radii of simulated clusters are defined \citep{Lahen2025a}. We note that at increased metallicity (10\% solar) the trend of model size evolution remains similar (N. Lahén, priv. comm.). Although conditions in the \citet{Lahen2025a} models and in \object{M\,31} differ, making direct comparisons and drawing conclusions difficult, the general similarity between observations and the state-of-the-art modelling results is nevertheless encouraging.

\citet{Lahen2025a} find that $\tt SPHGAL + KETJU$ clusters become exposed within the first few Myr, once the first photoionising stars ignite and their pre-SN stellar winds remove the remaining natal gas. This is supported by recent studies of young embedded SCs in nearby galaxies \citep{Knutas2025, Pedrini2026}, which show that clusters become fully exposed within $\sim$6\,Myr. As a result of this early gas expulsion, low-mass SCs experience substantial mass loss, significantly reducing their gravitational potential. This is followed by subsequent mass loss driven by stellar evolution and dynamical processes, leading to a rapid cluster expansion immediately after the ignition of massive stars. 

The similarity of the evolutionary trends between our observational results and the $\tt SPHGAL + KETJU$ models suggests that a similar mechanism – driven by internal cluster processes such as mass loss due to stellar feedback, stellar evolution, and internal gravitational interactions – may be responsible for the early structural evolution of SCs in the \object{M\,31} disc. Additionally, N-body simulations of star clusters in which the residual star-forming gas has a shallower density profile than the embedded stellar component show that, within the first 20\,Myr after gas expulsion, clusters can lose a large fraction of their mass and expand substantially; after $\sim$20\,Myr the surviving bound part of the cluster ceases to expand \citep{Shukirgaliyev2017}. This provides further support for interpreting our observational results as evidence of a similar phase of rapid early cluster expansion. Furthermore, \citet{Jorgensen2025}, who simulated the impact of giant molecular clouds (GMCs) on the evolution of young SCs in \object{M\,51}, found that such effects are not detectable over a timescale of 200\,Myr. This further supports our interpretation that the large increase of $R_{50}$ with age during the first $\sim$20\,Myr observed in Fig.\,\ref{fig2}b mainly reflects ongoing internal cluster processes rather than the effects of stochastic external perturbations, such as GMC encounters.

\begin{figure}
	\centering
	\includegraphics[width=9cm]{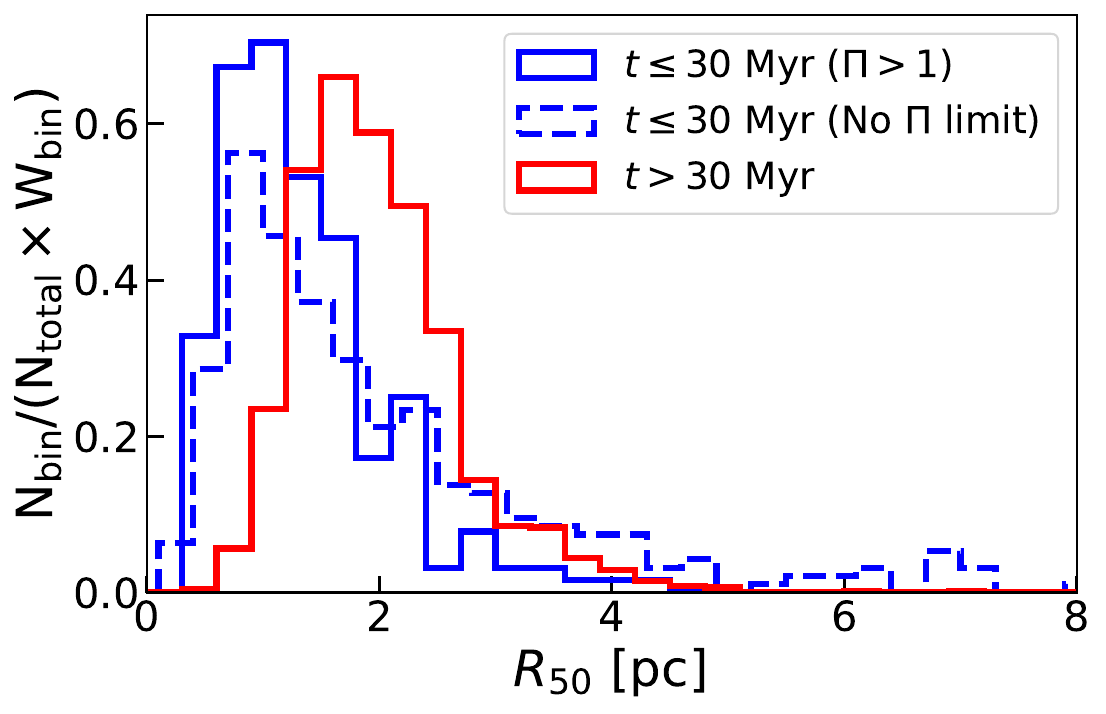}
	\caption{Comparison of cluster half-light radii distributions for \object{M\,31} clusters younger than (blue line) and older than (red line) 30\,Myr. The dashed histogram shows the distribution of all younger clusters; solid line histograms show the distributions after applying the $\Pi>1$ selection criterion.}
	\label{fig3}
\end{figure}

\begin{figure}
	\centering
	\includegraphics[width=9cm]{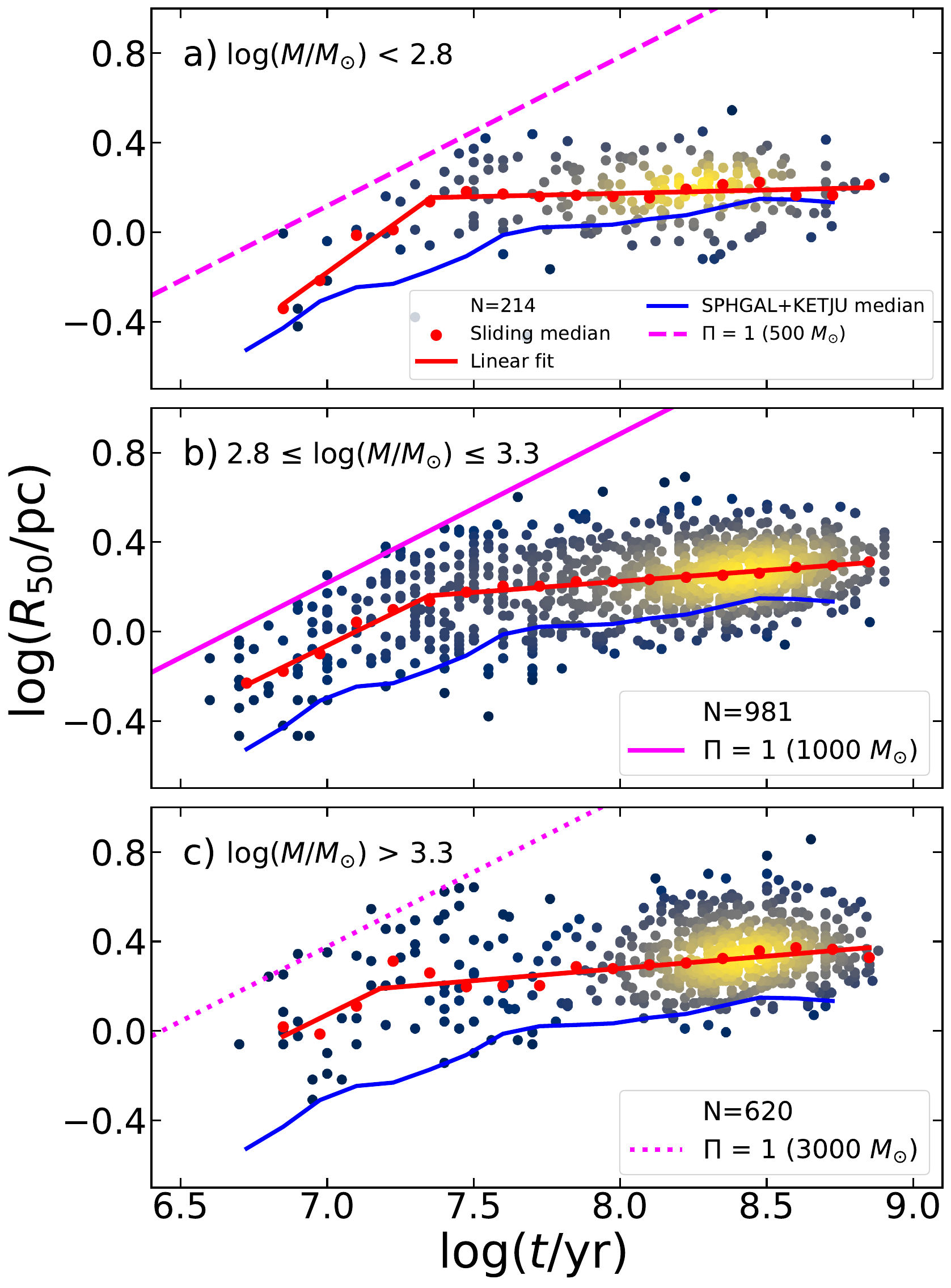}
	\caption{\object{M\,31} disc cluster half-light radii, $R_{50}$, versus their ages. Each panel shows a different cluster mass range (indicated in panels). The $\Pi>1$ criterion is applied in all panels. Median binning and linear fitting ranges are the same as in Fig.\,\ref{fig2}b.}
	\label{fig4}
\end{figure}

Alternatively, the observed trend of increasing $R_{50}$ with age early on could be driven by selection effects. In particular, the adoption of a $\Pi$ threshold may introduce a bias. As shown in Fig.\,\ref{fig2}, during the first $\sim$30\,Myr the slope of the $\Pi>1$ threshold excludes a substantial number of the youngest and most extended objects, introducing the possibility that the apparent growth is artificially induced by such selection. This effect becomes more evident when a stricter threshold (for example, $\Pi>2$) is applied. However, we note that very compact objects ($R_{50}\approx 0.5-1.0$\,pc, corresponding to $\sim$2.5–5.0 ACS/WFC pixels) are present among clusters younger than $\sim$100\,Myr but absent at older ages. We argue that SCs of this size in the 100–300\,Myr age range should not be strongly affected by selection effects: they remain relatively bright and blue compared to the average sky background which is dominated by older stellar populations in the \object{M\,31} disc, and are extended enough to be distinguished from point sources \citep{Daugevicius2024, Daugevicius2025}.

To further illustrate that cluster sizes change significantly during the first tens of Myr in the \object{M\,31} disc, in Fig.\,\ref{fig3}, we compare $R_{50}$ distributions of clusters younger than 30\,Myr (blue line) with those older than 30\,Myr (red line). A clear distinction is evident between the $R_{50}$ distributions of younger and older clusters. The peak $R_{50}$ values are $\sim$1.0\,pc for the younger population and $\sim$1.7\,pc for the older one. We note that even without limiting the sample by $\Pi$ value, the distribution of younger clusters remains generally similar, as shown by the blue dashed histogram in Fig.\,\ref{fig3}; all clusters older than 30\,Myr satisfy the $\Pi$>1 criterion. We interpret this as a representation of genuine physical cluster expansion, rather than an artefact introduced by selection effects. We emphasise that the duration of this early expansion phase ($\sim$20--30\,Myr) coincides with the lifetimes of SN progenitors, suggesting that stellar evolution processes, such as SN feedback and pre-SN winds of massive stars, may play a key role in this observed early cluster size growth.

We also examine how the cluster size–age relation depends on cluster mass. Figure\,\ref{fig4} shows $R_{50}$ as a function of cluster age for three mass intervals, arranged from the lowest masses in the top panel to the highest in the bottom panel. We find that the early increase in the characteristic $R_{50}$ with age is present for all three mass bins shown. The median trend of $R_{50}$ evolution has a similar shape to that of the $\tt SPHGAL+KETJU$ models in all three mass intervals; however, clusters with $2.8\leq \mathrm{log}(M/M_{\sun})\leq3.3$ show the most remarkable similarity with the shape of model tracks. For each mass interval, we derive the slope of the age--radius relation beyond $\sim$20\,Myr. For the lowest-mass clusters, the relation is virtually absent, a power-law index $\alpha=0.03\pm0.03$. Because the number of clusters older than 200\,Myr is small in this mass range, it is difficult to determine whether the slope is genuinely shallower for low-mass SCs than for the full sample. For the intermediate-mass clusters we obtain $\alpha=0.12\pm0.02$. The relation is slightly shallower for the highest-mass SCs, with $\alpha=0.08\pm0.02$, although this value is consistent within the uncertainties with both the intermediate-mass sample and the overall relation. We note that, at all ages, more massive clusters exhibit slightly larger median $R_{50}$ values. Indeed, a study by \citet{Pfalzner2016} found that in the case of embedded clusters in the solar neighbourhood, more massive ones tend to possess larger initial radii. A dependence of cluster size on mass is also suggested by theoretical models \citep{Gieles2016, Choksi2021, Grudic2021, Grudic2023}. The cluster mass – radius relation is discussed further in Sec.\,\ref{Sec4}. 

On the other hand, the fact that more massive clusters are slightly larger already at young ages could reflect their faster initial expansion. This might seem counterintuitive, as one would expect more massive clusters to be more tightly bound, retain their gas longer, and thus preserve their initial structure. However, recent studies \citep{Knutas2025, Pedrini2026} report that in nearby galaxies the emergence timescale from the embedded to the fully exposed phase is indeed shorter for more massive clusters. This implies that more massive clusters lose a significant fraction of their mass early on and can begin dynamical evolution sooner. Furthermore, as Fig.\,\ref{fig4} shows, clusters of different masses have similar radii, meaning that the more massive clusters are denser. Higher density leads to shorter dynamical timescales, $t_{\rm dyn}\propto\rho^{-1/2}$, and thus faster structural evolution. Additionally, the relation between cluster mass and the mass of the most massive member star \citep{Weidner2006, Weidner2010, Weidner2013, Zhou2026} implies that more massive clusters host higher-mass stars with shorter lifetimes. This results in an earlier onset of SN feedback and, more generally, a larger number of massive stars contributing stellar winds and SNe. Therefore, if early cluster evolution is primarily driven by stellar evolutionary processes, higher-mass clusters may evolve more rapidly.

\begin{figure}
	\centering
	\includegraphics[width=9.0cm]{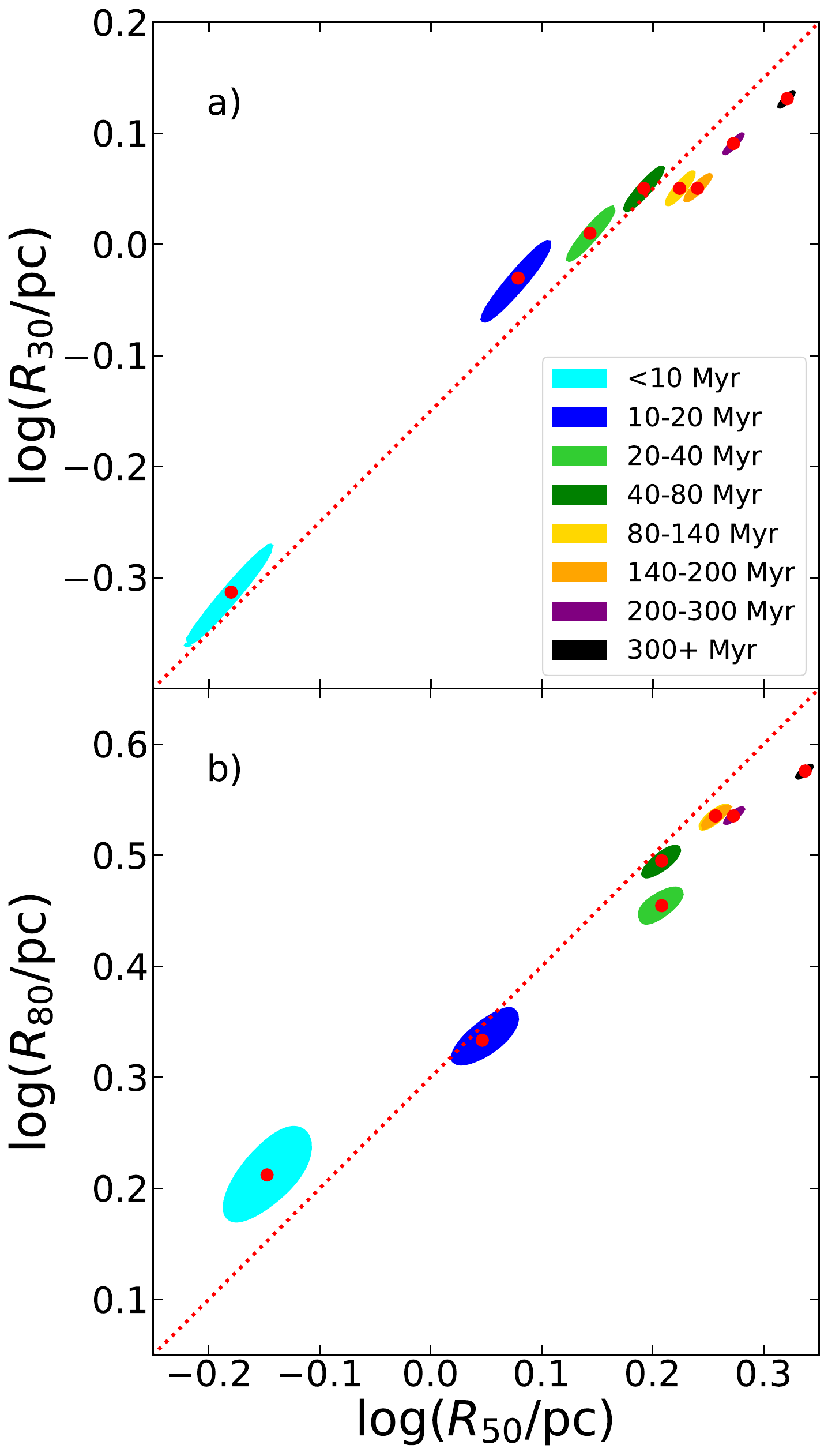}
	\caption{Evolution of star cluster structural parameters in the \object{M\,31} disc. Coloured patches show the 1$\sigma$ uncertainty of the distribution centres (marked by red dots) for each age bin. Centres and their uncertainties are derived from 3$\sigma$-clipped cluster samples satisfying $\Pi>1$. We note that in b) panel, centroids and their uncertainty regions of 80--140\,Myr and 140--200\,Myr age bins overlap. The red dotted line marks the diagonal along which parameters on both axes change the same number of times.}
	\label{fig5}
\end{figure}

\begin{figure}
	\centering
	\includegraphics[width=9cm]{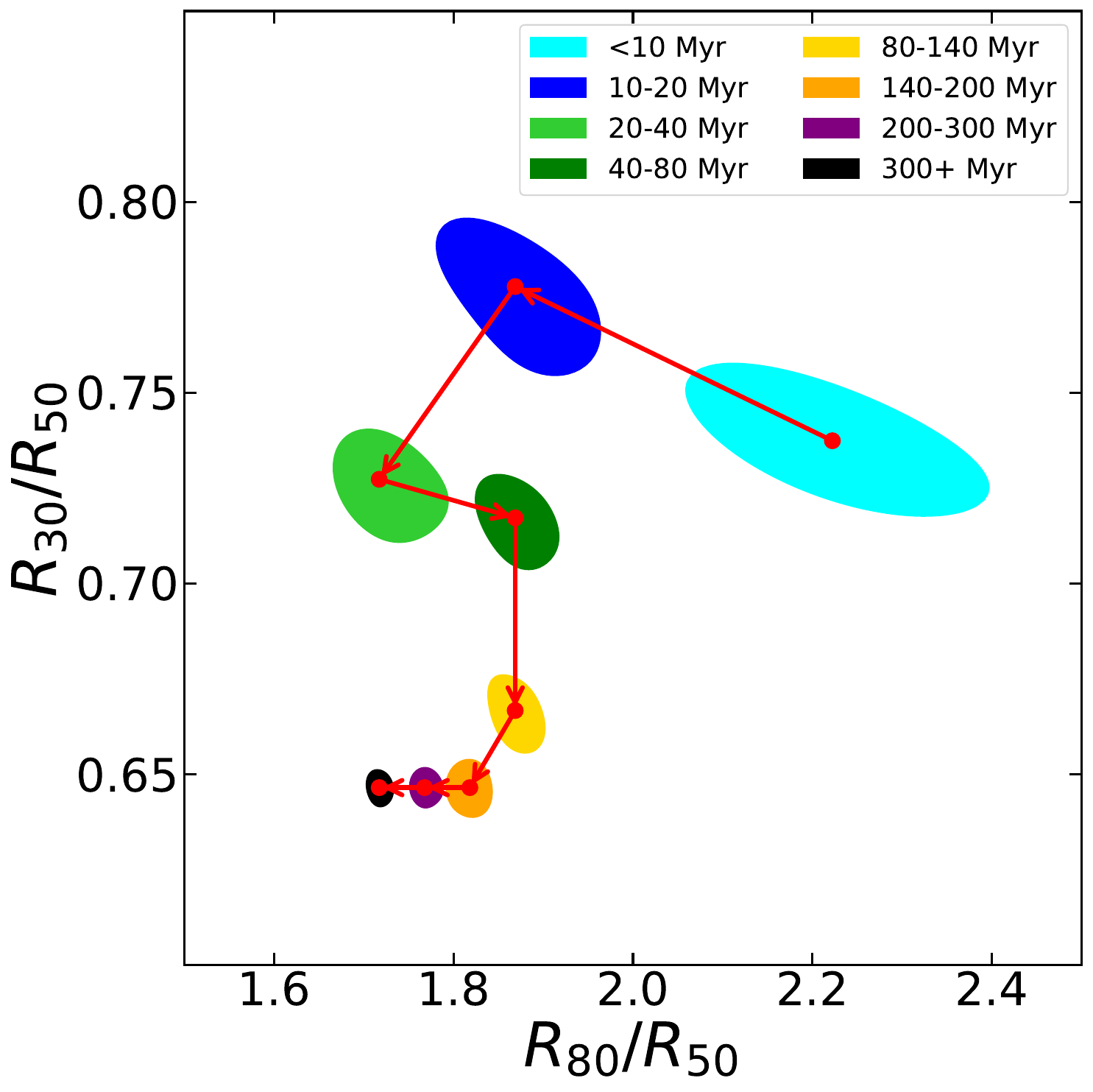}
	\caption{Evolution of the $R_{30}/R_{50}$ and $R_{80}/R_{50}$ ratios of star clusters in the \object{M\,31} disc. Coloured patches show the 1$\sigma$ uncertainties of the distribution centres (marked by red dots). Subsequent distribution centres are connected by red arrows. The centres and their uncertainties are derived from 3$\sigma$-clipped cluster samples satisfying $\Pi>1$.}
	\label{fig6}
\end{figure}

We further examine the evolution of $R_{30}$ and $R_{80}$ alongside $R_{50}$. Figure\,\ref{fig5} shows the age dependence of all three structural parameters. Distribution centres for each age bin are marked by red dots, while coloured patches indicate the 1$\sigma$ uncertainties of the distribution centres, derived from kernel density estimates (KDE). The centres and their uncertainties are computed from 3$\sigma$-clipped samples to remove outliers.

Because the parameters are correlated, we perform two-dimensional $\sigma$-clipping using the Mahalanobis distance \citep{Mahalanobis1936}. For a given pair of parameters (log($R_{50}$) and log($R_{30}$)), we estimate the mean vector $\mathbf{\bar{d}}$ and covariance matrix $\mathbf{C}$, and compute the squared Mahalanobis distance: 
\begin{fleqn}
	\begin{align}
	D^{2} = (\mathbf{d}-\mathbf{\bar{d}})^{\mathrm T}\mathbf{C}^{-1}(\mathbf{d}-\mathbf{\bar{d}}),
	\end{align}
\end{fleqn}
where $\mathbf{d}$ is the data vector. Objects with $D>3$ (correlated 3$\sigma$) are excluded. This method clips within an ellipse aligned with the intrinsic covariance, fully accounting for correlations. We adopt this procedure of $\sigma$-clipping throughout the remainder of this paper.

In Fig.\,\ref{fig5}, we see that all three structural parameters increase with age. During the first $\sim$40\,Myr, this growth is gradual; while $R_{30}$ and $R_{50}$ appear to increase at a similar rate, $R_{80}$ evolves more slowly. The evolution becomes more irregular in the $\sim$(40–200)\,Myr range, where one parameter may remain nearly constant while the others continue to increase. After this period, gradual expansion continues.

Figure\,\ref{fig6} shows the $R_{30}/R_{50}$ versus $R_{80}/R_{50}$ space, tracing the relative evolution of outer and inner cluster regions. During the first $\sim$20\,Myr, $R_{30}/R_{50}$ remains nearly unchanged while $R_{80}/R_{50}$ decreases, confirming that $R_{80}$ grows more slowly than $R_{30}$ and $R_{50}$. Between $\sim$20\,Myr and 140\,Myr, $R_{80}/R_{50}$ stays roughly constant, whereas $R_{30}/R_{50}$ decreases significantly, consistent with the nearly constant $R_{30}$ seen in Fig.\,\ref{fig5}a. This period coincides with the lifetimes of massive stars, suggesting that the observed structural evolution may be primarily linked to stellar-evolution-driven processes such as SN feedback and stellar winds.

In recent years, growing evidence for primordial mass segregation in young star clusters has emerged \citep{Plunkett2018, Nony2021, Zhang2022, Xu2024}, with more massive stars preferentially concentrated towards cluster centres. If such segregation is strong and expansion is driven by stellar-evolution-induced mass loss, a more pronounced core expansion would be expected, with $R_{30}$ increasing faster than $R_{50}$ and $R_{80}$. Although $R_{30}$ and $R_{50}$ indeed grow faster than $R_{80}$ during the first $\sim$40\,Myr (Fig.\,\ref{fig5}), we do not observe a clear difference when comparing $R_{30}$ and $R_{50}$. Consequently, it is difficult to confidently assess whether the observed trends reflect the presence of initial mass segregation. Beyond $\sim$140\,Myr, $R_{30}/R_{50}$ becomes stable and constant, while $R_{80}/R_{50}$ steadily decreases, indicating a slower expansion of outer regions. 

Such structural evolution may suggest that clusters begin to approach a tidally limited regime: as the outer layers expand towards a tidal boundary defined by the galactic tidal field, further growth of $R_{80}$ is suppressed, while inner cluster regions continue to expand. Although \citet{Heggie2003} suggest that the effective radius of tidally limited clusters undergoing mass loss should contract while the core expands, the continued increase of $R_{50}$ implies that \object{M\,31} disc clusters are unlikely to be fully tidally limited and instead evolve as isolated systems. However, structural parameters derived from aperture photometry are subject to stochastic effects \citep{Daugevicius2024} and thus caution is required when interpreting these results. The trends shown in Figs.\,\ref{fig5} and \ref{fig6} suggest that star clusters in the \object{M\,31} disc undergo rather complex structural evolution.

\begin{figure*}
	\centering
	\includegraphics[width=18.0cm]{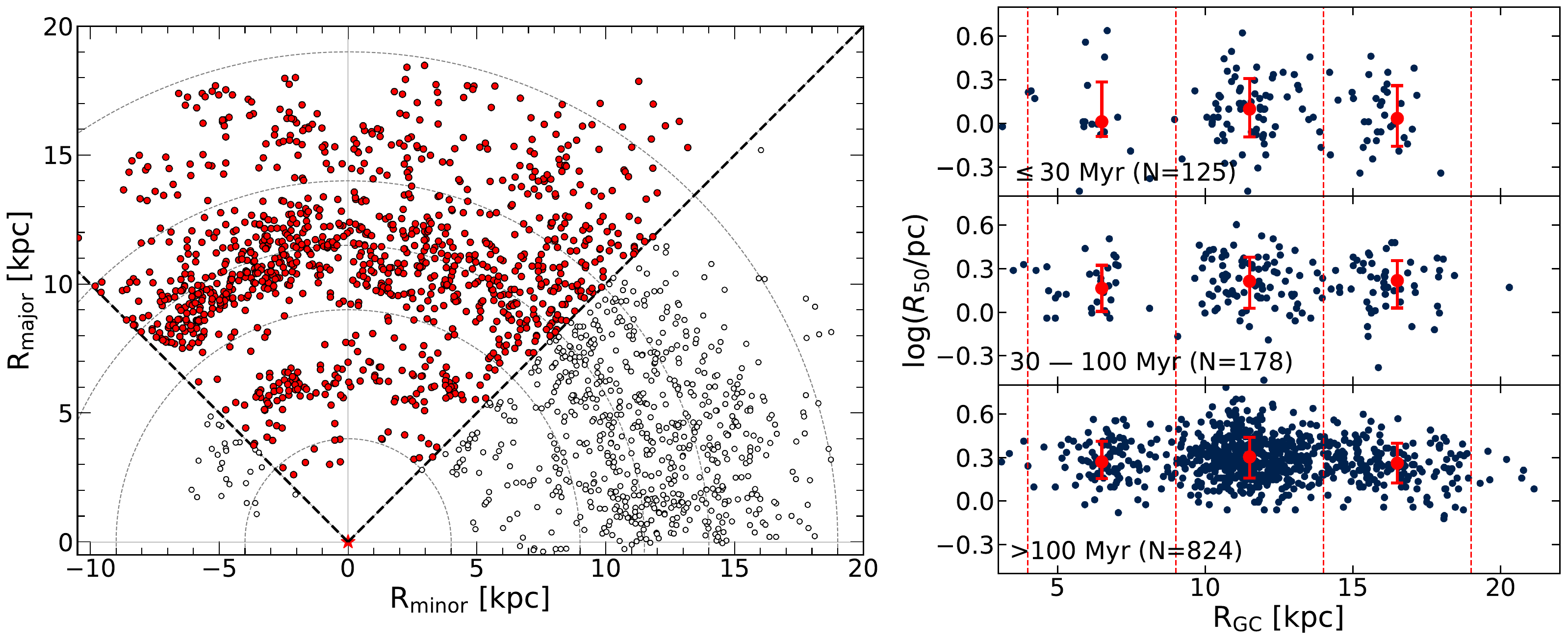}
	\caption{Dependence of cluster half-light radius ($R_{50}$) on galactocentric distance ($\rm R_{GC}$). We consider only clusters located near the major axis of the M31 disc, defined by an azimuthal angle relative to the major axis of –45\textdegree~ to +45\textdegree. Left: deprojected map of M31 disc star clusters. The azimuthal angle cut is indicated by black dashed lines; clusters within the cut are shown as filled red symbols, while those outside the cut are shown as open circles. Right: $R_{50}$ as a function of $\rm R_{GC}$ for three different age intervals (indicated in each panel). Red markers show the median $R_{50}$ in each $\rm R_{GC}$ bin (their boundaries are denoted by vertical red dashed lines), with error bars showing the 16th-84th percentile range. Only clusters satisfying $\Pi>1$ are included in both panels.}
	\label{fig7}
\end{figure*}

If a significant cluster age gradient with galactocentric distance, $\rm R_{GC}$, exists within the studied part of the \object{M\,31} disc, the apparent structural evolution with age discussed in this section could instead result from environmental effects – such as variations in the galactic tidal field or location-dependent cluster structure – rather than expansion driven by internal evolutionary processes. This possibility is particularly relevant given previous studies reporting the presence of an $R_{50}$--$\rm R_{GC}$ relation in some nearby spiral galaxies \citep{Scheepmaker2007, SanRoman2012}, as well as the recent study by \citet{Tang2026}, based on the LEGUS survey, which shows that the demography of young clusters depends on their environment. Therefore, we examine whether any trends between cluster size and $\rm R_{GC}$ are present to assess whether the apparent structural evolution is driven by internal or environmental mechanisms.

We explore this in Fig.\,\ref{fig7}. For this analysis, we restrict the sample to clusters located near the major axis, defined by azimuthal angles of –45\textdegree~ to +45\textdegree~ relative to it, as shown in the left panel of Fig.\,\ref{fig7}. This selection is motivated by the high inclination angle ($\sim$77.5\textdegree), which introduces increasing uncertainty in $\rm R_{GC}$ for clusters located farther from the major axis due to deprojection effects. This is illustrated in the same panel: at projected distances of $\sim$(10–15)\,kpc along the X-axis, beyond the black dashed line marking the +45\textdegree~direction, three distinct ring structures become increasingly difficult to distinguish as a result of these deprojection uncertainties.

The right panel of Fig.\,\ref{fig7} shows $R_{50}$ as a function of $\rm R_{GC}$ for three cluster age intervals. Overall, we detect no clear cluster size gradient with galactocentric distance in any age bin; however, a large scatter of $R_{50}$ values is observed at a fixed $\rm R_{GC}$. We note that the structural parameter ratios $R_{30}/R_{50}$ and $R_{50}/R_{80}$ behave similarly, showing no clear dependence on $\rm R_{GC}$. A notable feature emerges for clusters with $t<100$\,Myr: the three radial rings are rather clearly separated, whereas at older ages they are more blended. This may suggest that clusters migrate from their birthplaces on timescales longer than $\sim$100\,Myr, implying that SCs can serve as excellent tracers of previous star-forming regions over extended timescales.

In general, we find no evidence for any correlation between cluster sizes and their galactocentric distances, indicating that the tidal field gradient within the studied region of the \object{M\,31} disc does not significantly influence cluster structural properties. This supports our earlier suggestion that the structural evolution is more likely driven by internal cluster processes than by environmental effects.

In summary, we demonstrate that bound clusters in the disc of \object{M\,31} undergo rapid expansion during the first $\sim$20\,Myr, increasing their sizes by more than a factor of two. This behaviour is similar to the $\tt SPHGAL+KETJU$ cluster models \citep{Lahen2025a}, in which early residual gas expulsion leads to significant mass loss and triggers rapid expansion driven by stellar evolution and internal dynamics. Our results are also highly consistent with recent findings for young clusters in the \object{MW} \citep{DellaCroce2024}, which show, based on the kinematic \textit{Gaia} data, that Galactic SCs younger than 30\,Myr are undergoing rapid expansion. 

We further find that \object{M\,31} disc clusters show no correlation between their sizes and distances from the galactic centre, indicating that external tidal effects are likely to be insignificant in their early evolution. We therefore interpret the observed early structural evolution of the clusters as expansion of initially compact, not tidally limited clusters, driven mainly by stellar-evolution-induced mass loss and two-body relaxation, which happens after the early residual natal gas removal. These clusters are gradually evolving towards a tidally limited regime, eventually reaching virial equilibrium.

\section{Mass – radius relation}
\label{Sec4}

\begin{figure*}
	\centering
	\includegraphics[width=18.0cm]{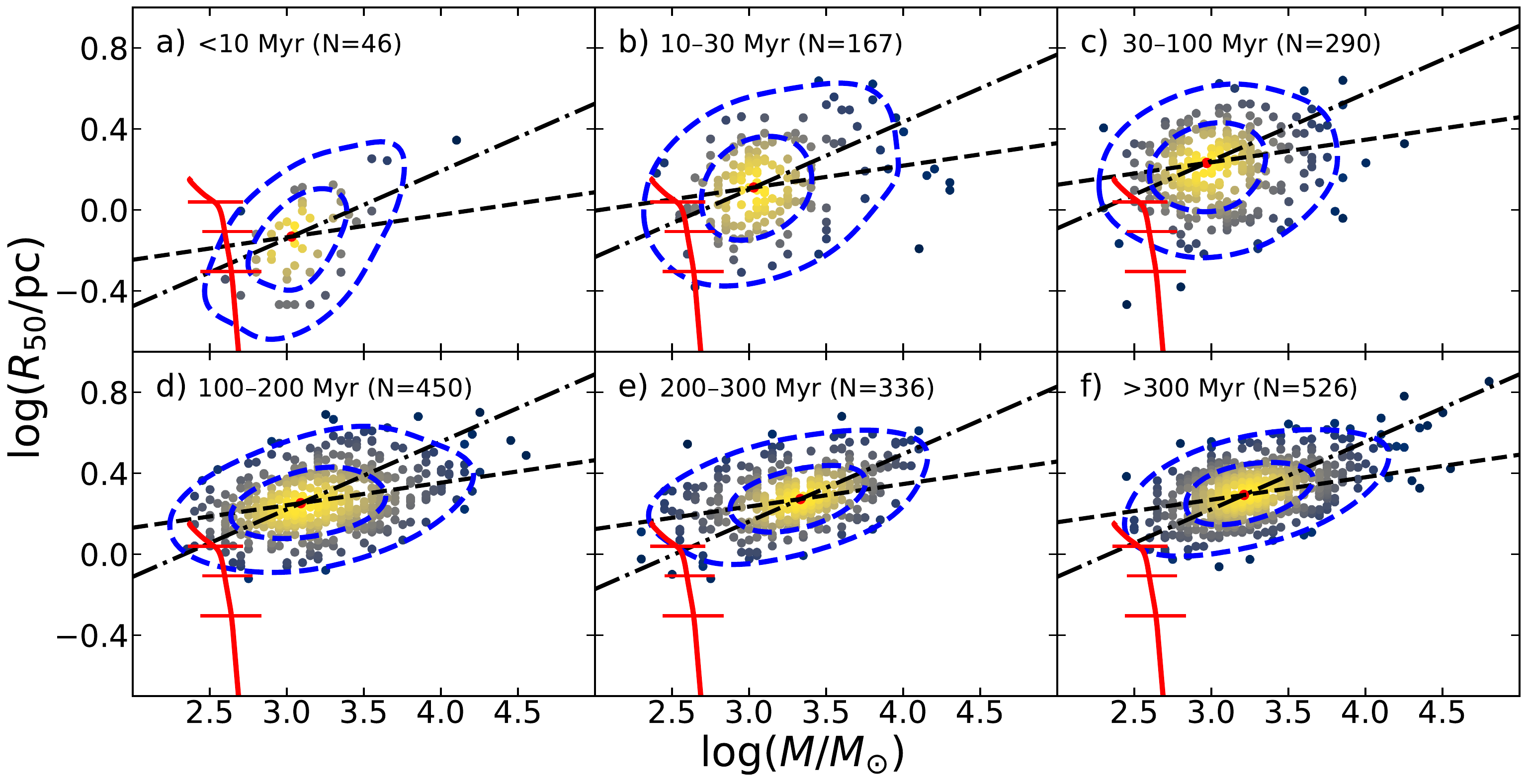}
	\caption{Evolution of the cluster mass-radius relation. Each panel shows $R_{50}$ as a function of mass for different age intervals (indicated in each panel). The numbers shown next to the age intervals indicate the sample sizes. The constant density relation, $R_{50} \propto M^{1/3}$ (black dash-dotted line), and the equilibrium relation, $R_{50} \propto M^{1/9}$ (black dashed line), are shown for reference. Blue dashed contours represent KDEs, with the inner contour enclosing 39.3\% of the KDE integral around the KDE peak (corresponding to a 1$\sigma$ region), while the outer one encloses 86.5\% (2$\sigma$ region). The red curve shows the median evolutionary track of the $\tt SPHGAL + KETJU$ models \citep{Lahen2025a}; horizontal bars at 10, 30, and 100\,Myr (from bottom to top) indicate the full range of model masses at these ages. In all panels, only clusters satisfying the $\Pi > 1$ criterion are included.
	}
	\label{fig8}
\end{figure*}

In this section, we examine the cluster mass–radius relations. Figure\,\ref{fig8} shows the distribution of SCs in the mass–$R_{50}$ space for six cluster age bins. For reference, we include the constant density relation, $R_{50} \propto M^{1/3}$ (black dash-dotted line), which is predicted by various analytic and numerical models \citep{Choksi2021, Grudic2021}, as well as the equilibrium relation, $R_{50} \propto M^{1/9}$ (black dashed line), derived from a model that combines the effects of tidal shocks and two-body relaxation \citep{Gieles2016, Renaud2018}. The blue dashed contours represent KDEs, with the inner contour denoting an area enclosing 39.3\% of the KDE integral around the KDE peak (corresponding to a 1$\sigma$ region), while the outer one encloses 86.5\% (2$\sigma$ region). The red curve shows the median evolutionary track of $\tt SPHGAL + KETJU$ models, with horizontal bars indicating the full range of model masses at 10, 30, and 100\,Myr.

Panels a–c) of Fig.\,\ref{fig8} clearly demonstrate the early cluster expansion discussed in Sec.\,\ref{Sec3}, as the cluster distribution shifts upward along the log($R_{50}/\rm pc$) with increasing age during the first few tens of Myr, similar to the behaviour shown by evolutionary tracks of $\tt SPHGAL + KETJU$ models. Although visual inspection of the KDE contours in the youngest bins (<10\,Myr and 10–30\,Myr) may suggest a steep trend – possibly even steeper than $R_{50}\propto M^{1/3}$ – these bins are strongly affected by the adopted $\Pi>1$ selection criterion and may therefore be subject to selection effects. While the analytic model of \citet{Choksi2021} predicts a steep initial relation, $R_{50}\propto M^{1/2}$, which later approaches $R_{50}\propto M^{1/3}$, we cannot confirm such a trend, especially considering that these young clusters are likely undergoing rapid expansion and each may be at different evolutionary stages. Consequently, we are unable to claim the presence of a pronounced mass–radius relation for clusters younger than 100~Myr (see Fig.\,\ref{fig8}a–c).

However, a shallow yet significant mass–radius relation emerges for clusters older than $\sim$100\,Myr (see Fig.\,\ref{fig8}d–f), once the apparent rapid cluster structural evolution settles. Visually, the slope appears similar across all subsequent age bins beyond 100\,Myr, and lies between the $R_{50}\propto M^{1/3}$ and $R_{50}\propto M^{1/9}$ scaling relations.

\begin{figure*}
	\centering
	\includegraphics[width=18.0cm]{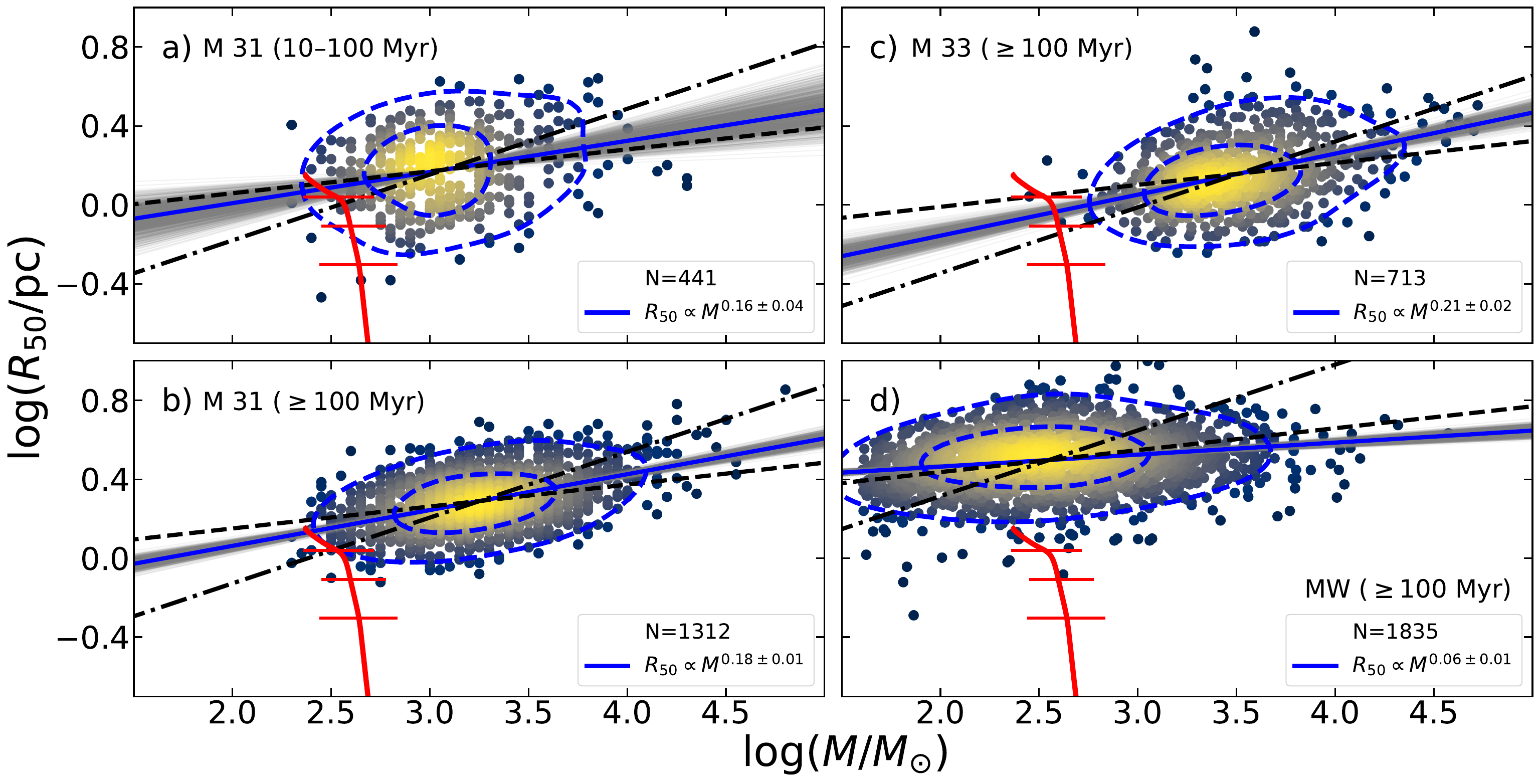}
	\caption{Star cluster mass–radius relations. Panel a) shows \object{M\,31} clusters with ages 10–100\,Myr; panel b) – \object{M\,31} clusters older than 100\,Myr. Panel c) presents \object{M\,33} clusters older than 100\,Myr, based on the PHATTER star cluster catalogue \citep{Johnson2022}. Panel d) shows \object{MW} clusters older than 100\,Myr, based on the \citet{Hunt2024} \textit{Gaia} cluster catalogue (only objects classified as ``open clusters'' are presented). The numbers shown in the legends indicate the sample sizes. The blue solid line indicates the slope of the major axis of the covariance (confidence) ellipse (see text); the numerical values of the slopes are given in the legends as the power-law indices of cluster mass–radius relations. Grey lines show individual slopes obtained from 1000 bootstrap resampling iterations. The constant density relation, $R_{50} \propto M^{1/3}$ (black dash-dotted line), and the equilibrium relation, $R_{50} \propto M^{1/9}$ (black dashed line), are shown for reference. The inner blue dashed contour encloses 39.3\% (1$\sigma$ region) of the KDE integral around the KDE peak, while the outer one encloses 86.5\% (2$\sigma$ region). The red curve and horizontal bars show the median $\tt SPHGAL + KETJU$ model track, as in Fig.\,\ref{fig8}. In all panels, only clusters satisfying the $\Pi>1$ criterion are included.}
	
	\label{fig9}
\end{figure*}

In Fig.\,\ref{fig9}, we examine the mass–radius relation for \object{M\,31} disc clusters more quantitatively, separating the clusters into two age groups of 10–100\,Myr and $>$100\,Myr. In both cases, cluster distribution in the log($R_{50}/\rm pc$) versus log($M/M_{\sun}$) space is reasonably well approximated by a two-dimensional normal distribution. We therefore characterise it using covariance (or confidence) ellipses derived from the covariance matrix. We define the power-law index of the mass–radius relation as the slope of the covariance ellipse's major axis. To estimate the slope and its uncertainty, we perform 1000 bootstrap resampling iterations, deriving a slope for each case (grey lines). The final power-law index is taken as the median of these individual slopes (blue solid line), with the uncertainty defined as half the 16th-84th percentile range, which we find to be symmetrical.

For the young (10–100\,Myr) clusters, we derive a shallow relation, $R_{50}\propto M^{0.16\pm0.04}$. However, the uncertainty of the power-law index is relatively large (Fig.\,\ref{fig9}a). For clusters older than 100\,Myr, we obtain a more robust relation, $R_{50}\propto M^{0.18\pm0.01}$ (Fig.\,\ref{fig9}b). Moreover, the $\Pi>1$ criterion does not affect this age range, as no clusters are rejected by its adoption; therefore, no related selection effects are expected. We note that older ($t\geq100$\,Myr) cluster sample consists of three age bins with slightly increasing slopes: 0.15$\pm$0.02 for 100--200\,Myr, 0.18$\pm$0.02 for 200--300\,Myr, and 0.20$\pm$0.02 for $>$300\,Myr. While this behaviour could hint at a possible weak steepening with age of the mass--radius relation beyond 100\,Myr, we are unable to draw a definite conclusion that such a trend exists. The slight variation in slope could also be influenced by selection effects affecting low-mass clusters ($\rm M\lesssim1000\,M_{\odot}$) at ages beyond 300\,Myr \citep{Daugevicius2025}. Nonetheless, the main result remains unchanged: clusters older than 100\,Myr in the M~31 disc exhibit a shallow mass--radius relation.

The derived slope is similar to that found in \object{M\,51} by \citet{Hwang2010} (0.16$\pm$0.02), but shallower than the relations reported for individual galaxies outside the Local Group or for SC samples combined from multiple nearby galaxies \citep[$R_{50}\propto M^{0.25-0.33}$;][]{Ryon2015, Krumholz2019, Brown2021}. It is steeper than the \citet{Gieles2016} equilibrium relation ($R_{50}\propto M^{1/9}$), yet significantly shallower than the $R_{50}\propto M^{1/2}$ \citep{Choksi2021}, or $R_{50}\propto M^{1/3}$ \citep{Grudic2021}, and slightly shallower than $R_{50}\propto M^{1/4}$ \citep{Grudic2023} scaling relations.

Finally, we compare the \object{M\,31} disc results with other Local Group galaxies. In the right panels of Fig.\,\ref{fig9}, we present mass–radius relations for clusters with $t\geq100$\,Myr, derived using the same methodology in \object{M\,33} (Fig.\,\ref{fig9}c) and \object{MW} (Fig.\,\ref{fig9}d), based on the cluster catalogues by \citet{Johnson2022} and \citet{Hunt2024}, respectively. 

In \object{M\,33}, we derive a somewhat steeper but broadly similar relation ($R_{50}\propto M^{0.21\pm0.02}$) compared to \object{M\,31}. We note that the \citet{Johnson2022} \object{M\,33} catalogue lacks clusters with $M\leq1000~M_{\sun}$ compared to our SCs sample, which may partly account for this difference. 

In contrast, the \object{MW} shows a significantly shallower relation, $R_{50}\propto M^{0.06\pm0.01}$, implying that cluster size is virtually independent of mass. This trend is consistent with results reported for some galaxies beyond the Local Group, where very shallow or absent mass–radius relations are found \citep{Zepf1999, Larsen2004, Bastian2005, Scheepmaker2007, Bastian2012, Ryon2017}. Within the Local Group, \citet{Gatto2021} examined the core radius, $r_{c}$, and its correlation with cluster mass in the \object{SMC}. They derived the mass–core radius relation $r_{c}\propto M^{0.34}$, consistent with the $M^{1/3}$ scaling \citep{Krumholz2019, Choksi2021, Grudic2021}.

Although we cannot draw definite conclusions based on the comparison of \object{M\,31} with \object{M\,33} and \object{MW} (Fig.\,\ref{fig9}) due to differences in instrumentation, cluster parameter determination methods, and sample sizes, the results suggest that the cluster mass–radius relation varies substantially within the Local Group. Combined with the diverse relations reported for more distant galaxies \citep{Zepf1999, Larsen2004, Bastian2005, Hwang2010, Bastian2012, Ryon2015, Ryon2017, Brown2021}, this indicates that, despite the lack of significant environmental variation within the disc of \object{M\,31}, the mass–radius relation likely depends strongly on global galactic environments. Such a variety of scaling values may be driven by differences in pressure, natal gas density, metallicity, tidal field strength, turbulence, the frequency of tidal shocks, and other physical processes governing star formation and cluster evolution. Thus, we emphasise the need for further theoretical work to explain these contrasting results among the discs of different galaxies.

\section{Summary and conclusions}
\label{Sec5}

We have investigated the structural evolution of star clusters and the correlations between their sizes and other properties (age, mass, and galactocentric distance) in the disc of \object{M\,31}. We used cluster parameters derived based on the high resolution HST ACS+WFC3 data from the PHAT survey \citep{Dalcanton2012, Johnson2015}. The analysed cluster sample comprises 1918 objects younger than 800\,Myr.

We find that young clusters in the north-eastern part of the \object{M\,31} disc are predominantly compact, with a median half-light radius of $R_{50}\approx1.8$\,pc. This is consistent with earlier results for the south-western region of the \object{M\,31} disc \citep{Sableviciute2006, Sableviciute2007, Vansevicius2009}, indicating that clusters are compact across the entire \object{M\,31} disc.

Furthermore, we show that bound clusters undergo rapid expansion within the first $\sim$20\,Myr, with $R_{50}$ increasing by more than a factor of two. We interpret the observed structural evolution as follows: after the early removal of residual natal gas and associated mass loss, initially compact clusters undergo rapid expansion due to internal cluster processes, such as stellar-evolution-driven mass loss and two-body relaxation, gradually evolving towards a tidally limited virialised state. Importantly, these results are in good agreement with recent findings for young clusters in the \object{MW} by \citet{DellaCroce2024}, who report rapid and violent cluster expansion on a similar $\sim$30\,Myr timescale based on \textit{Gaia} kinematics. Beyond 20\,Myr, \object{M\,31} clusters follow a shallow age--radius relation, $R_{50}\propto t^{0.12\pm0.01}$.

We detect no significant dependence of $R_{50}$ on galactocentric distance, suggesting that the majority of young \object{M\,31} disc clusters are not yet tidally limited, implying that the galactic field plays a minor role in their early evolution. This further supports our conclusion that the observed early cluster evolution in \object{M\,31} is driven by internal cluster processes. In addition, our results suggest inefficient radial migration of clusters over $\sim$100\,Myr or possibly even longer timescales, implying that clusters can provide key insight into past star-forming regions over extended periods of time.

We find a rather weak mass–radius relation for clusters younger than 100\,Myr. In contrast, clusters older than 100\,Myr exhibit a clear relation, $R_{50}\propto M^{0.18\pm0.01}$, with no significant radial dependence. However, this slope, together with the observed mass–radius relations we find for \object{M\,33} and \object{MW} clusters and those reported in more distant galaxies \citep{Zepf1999, Larsen2004, Bastian2005, Hwang2010, Bastian2012, Ryon2015, Ryon2017, Brown2021}, highlights the need for further theoretical work to consistently explain the wide range of slope variations, and to better understand cluster formation and evolution across different galactic disc environments.

\begin{acknowledgements}
We are grateful to Natalia Lahén for kindly sharing the $\tt SPHGAL+KETJU$ model data and for providing helpful comments and suggestions, which were essential for the analysis and discussion presented in this paper. We also thank Emily L. Hunt for providing clarifications regarding the \citet{Hunt2024} catalogue data and for additional comments and suggestions that were valuable for this study. This research has made use of: the NASA/IPAC Extragalactic Database (NED), which is funded by the National Aeronautics and Space Administration and operated by the California Institute of Technology; SAOImage DS9, developed by the Smithsonian Astrophysical Observatory (\url{https://ds9.si.edu}); Astropy (\url{http://www.astropy.org}), a community-developed core $\tt Python$ package for Astronomy \citep{Astropy2013, Astropy2018}; $\tt APLpy$, an open-source plotting package for $\tt Python$ \citep{aplpy2012}; SciPy, an open-source scientific library for $\tt Python$ \citep{scipy2020}; Matplotlib, a plotting library for $\tt Python$ \citep{Hunter2007}; TOPCAT (\url{https://www.star.bristol.ac.uk/mbt/topcat}). It is based on observations made with the NASA/ESA \textit{Hubble} Space Telescope, and obtained from the \textit{Hubble} Legacy Archive, which is a collaboration between the Space Telescope Science Institute (STScI/NASA), the Space Telescope European Coordinating Facility (ST-ECF/ESA) and the Canadian Astronomy Data Centre (CADC/NRC/CSA). This project has received financial support from the Research Council of Lithuania (LMTLT), agreement No S-MIP-24-98. EK acknowledges financial support from FTMC under agreement No 3300-BL15. Computations were performed on the supercomputer GALAX of the Center for Physical Sciences and Technology, Lithuania. 
\end{acknowledgements}

\bibliographystyle{aa}
\bibliography{aa62808-26}

\end{document}